\documentclass[epjCONF,columns]{svjour} 
\usepackage{graphics}
\usepackage[varg]{txfonts} 
\usepackage[latin1]{inputenc}
\session-title{Presented at the 2011 Hadron Collider Physics symposium (HCP-2011), Paris, France, November 14-18 2011}
\begin{document}
\title{Recent results from HERA and their impact for LHC}
\author{K. Lipka\inst{1}\fnmsep\thanks{\email{katerina.lipka@desy.de}} for the H1 and ZEUS Collaborations}
\institute{Deutsches Electronen Synchrotron DESY Hamburg}
\abstract{
Recent measurements of inclusive and semi-inclusive deep inelastic scattering in electron-proton collisions at HERA are reviewed.  
These measurements are used to determine the parton distribution functions (PDFs) of the proton, a necessary input to theory predictions
for hadron colliders. An introduction to the PDF determination with an emphasis on HERA PDFs is presented. Theory predictions based on HERAPDF 
are compared to a selection of recent LHC and Tevatron measurements. The impact of jet and charm production measurements on the PDFs is discussed.
} 
\maketitle
\section{Introduction}
\label{intro}
A deep understanding of the proton structure is one of the most important topics in modern particle physics. A precise knowledge of the 
Parton Distribution Functions (PDFs) of the proton is essential in order to make predictions for the Standard Model and beyond the 
Standard Model processes at hadron colliders.

The cross sections of processes in proton-(anti)proton collisions are factorized by a convolution of the matrix element of parton-parton interaction and the proton structure.
The latter is described by parton density functions (PDFs). A PDF, $f_i(x,Q^2)$, represents the probability of finding in the proton a parton $i$ (quark or gluon) carrying a fraction $x$ of the proton momentum with $Q$ being the energy scale of the hard interaction. In case of proton-(anti)proton interactions PDFs of both protons enter multiplicatively into the calculation of the process cross section. Therefore the precision of the PDFs is of particular importance for accurate cross-section predictions. 
In the last decades, the measurements of lepton-nucleon and proton-antiproton scattering have been used to determine the proton PDFs.  At low to medium $x$ the 
PDFs are constrained by HERA data. The measurements at fixed target experiments and Tevatron contribute mainly at high $x$.  The recent precise data from 
Tevatron and the LHC experiments have the potential to improve the precision on the PDFs further.
\section{Proton Structure and DIS at HERA}
\begin{figure}[h]
\center
\resizebox{0.5\columnwidth}{!}{\includegraphics{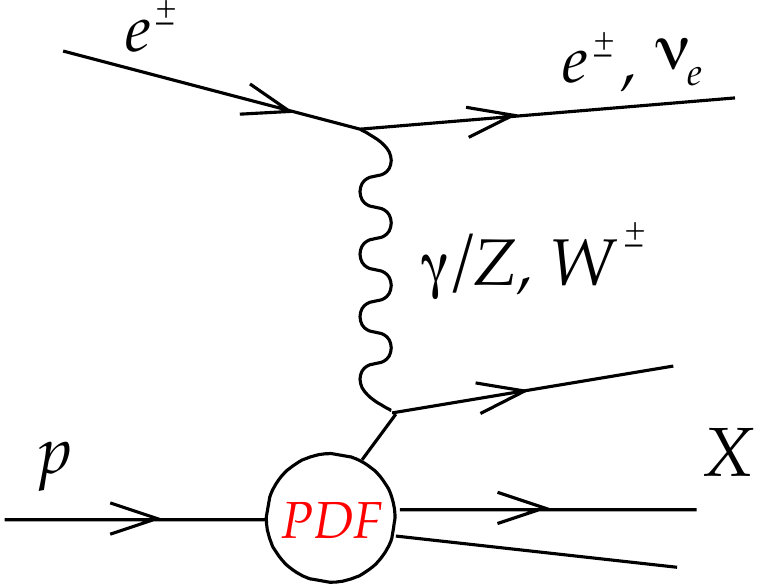} }
  \caption{\it Diagrams of neutral NC and charged CC current  deep inelastic scattering processes. The symbols denote the particles, the label "$X$" denotes 
 the hadronic final state.} 
 \label{dis_diagram}
\end{figure}

The knowledge of the proton PDFs is obtained to a large extent from the measurements of the structure functions in deep inelastic scattering (DIS) experiments. 
In Fig.~\ref{dis_diagram}. the diagram of DIS is represented. The lepton is scattered off  the nucleon via the exchange of a $\gamma$ or $Z^0$-boson (neutral current, NC, process) or via the exchange of a $W^{\pm}$-boson(charged current, CC). Here the scattering of an electron (or positron) off the proton is discussed.

The NC (and similarly CC) cross section can be expressed in terms of the generalized structure functions:
\begin{eqnarray}
  \nonumber
   \frac{d^2\sigma_{NC}^{e^{\pm} p}}{dxdQ^2}=\frac{2\pi\alpha^2}{xQ^4} 
     \big [ Y_{+} \tilde F_2^{\pm} \mp Y_{-}x \tilde F_3^{\pm} - y^2 \tilde F_L^{\pm} \big ],
\end{eqnarray}
where $Y_{\pm} = 1 \pm (1-y)^2$ with $y$ being the transferred fraction of the lepton energy. The (generalized) structure function $F_2$ ($\tilde F_2$)
is the dominant contribution to the cross section, $x \tilde F_3$ is important at high $Q^2$ and $\tilde F_L$ is sizable  only at high $y$. 
In the framework of perturbative QCD the structure functions are directly related to the parton distribution functions, i.e. in leading order (LO)  
$F_2$ is the momentum sum of quark and anti-quark distributions, 
$F_2 \approx x \sum e^2_q (q+ \overline q)$, and $xF_3$ is related to their difference, 
    $xF_3 \approx x \sum 2e_q a_q (q- \overline q)$. At higher orders, terms related to the gluon density distribution
($\alpha_s g$) appear.
In analogy to neutral currents, the inclusive CC $ep$ cross section can be expressed in terms of structure functions and in LO the $e^+p$ and $e^-p$ cross 
sections are sensitive to different quark densities:
\begin{eqnarray}
  \nonumber
    \begin{array}{rll}
   e^{+}:  & & \tilde \sigma_{CC}^{e^{+} p} = 
                x[\overline u +\overline c] + (1-y)^2 x[ d+s ]  \\
   e^{-}:  & & \tilde \sigma_{CC}^{e^{-} p} = 
                x[ u +c] + (1-y)^2 x[\overline d +\overline s ].
    \end{array}
\end{eqnarray}
 \vspace{0.07cm}

At HERA at DESY in Hamburg, electrons (or positrons) were collided with protons at centre-of-mass energies $\sqrt{s} = 225 - 318$~GeV. 
The measurements of the NC and CC cross sections from HERA extend the kinematic regime in $Q^2$ by more than two orders of magnitude with respect to the fixed target experiments and cover the wide $x$ range from $10^{-7}$ to 0.7. At the HERA collider experiments, H1 and ZEUS, the cross sections of  NC and CC DIS are measured with high precision. The measurements of the two experiments are combined and are further used to determine parton distribution functions HERAPDF~\cite{herapdf1.0} .  

\section{HERAPDF}
The PDFs are determined from the structure function measurements using the corresponding coefficient functions calculated to a certain order in perturbative QCD (pQCD). The structure functions, and in turn the PDFs, depend on $x$ and $Q$.  The $x$-dependence of the parton distributions is not yet calculable in pQCD and has to be parametrized at a certain starting scale $Q_0$. The dependence on $Q$ is described by the DGLAP evolution equations~\cite{dglap}. 
Starting from a parameterisation of the PDFs at a starting scale, either by making ad-hoc assumptions on their analytical form or by using the neural-net technology, fits to various sets of experimental data, with HERA DIS data being the backbone, are performed within the DGLAP evolution scheme. 
The resulting PDFs depend on the order in which the perturbative QCD calculation is performed, the assumptions about the PDF parametrization, the treatment of heavy quarks, the choice for the value of $\alpha_s (M_Z)$  and the treatment of the uncertainties. The data sets included in the PDF fit and the consistency of these data sets determines the experimental uncertainty of the PDFs. 
\begin{figure}[!h]
\center
\resizebox{0.75\columnwidth}{!}{\includegraphics{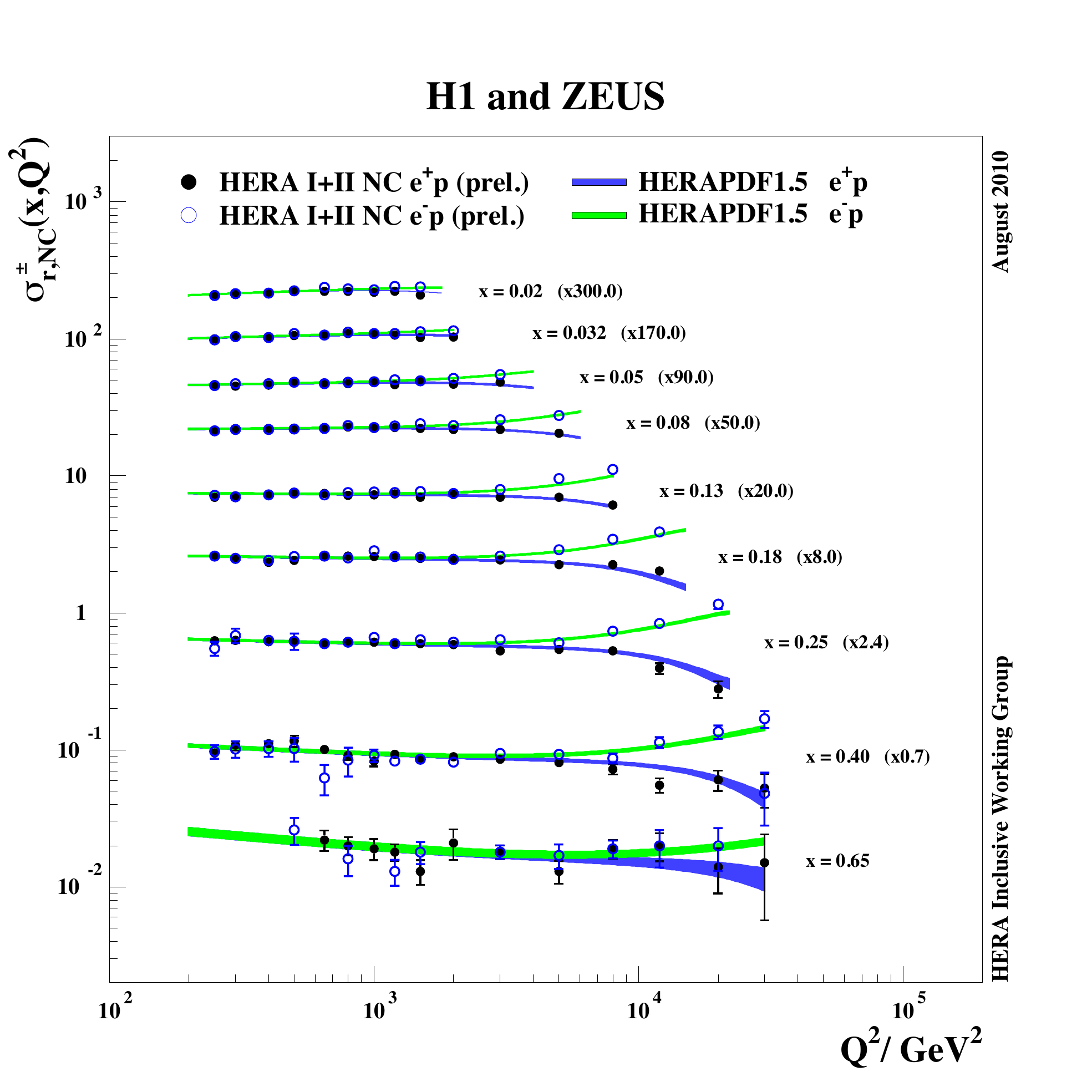}}
 \caption{\it Inclusive DIS cross sections for NC in $e^{\pm}$ collisions at HERA. The measurements of the H1 and ZEUS experiments are combined. Open (closed) symbols represent $e^{-}p$ ($e^{+}p$) scattering. The shaded curves represent QCD prediction based on HERAPDF1.5NLO.}
 \label{nc_hera15}
\end{figure}

The parton distributions HERAPDF ~\cite{herapdf1.0} are determined using only combined HERA DIS data, where the correlations of the systematic uncertainties are properly taken into account. This allows the usage of the conventional $\chi^2$ tolerance of $\Delta \chi^2=1$. Since this QCD analysis is solely based on $ep$ data, the PDFs do not depend on the approach for nuclear corrections needed for fixed target data. Several phenomenological schemes of heavy quark treatment can be used in the HERAPDF approach. Therefore direct tests of the models are possible. 
The full statistics of the HERA inclusive CC and NC data are used for NLO and NNLO QCD fits resulting in HERAPDF1.5~\cite{herapdf15}. As an example, the combined NC  cross sections are shown in Fig.~\ref{nc_hera15} together with QCD prediction based on HERAPDF1.5NLO.

The QCD analysis HERAPDF1.5 follows the formalism, model and paramatrisation assumptions as reported in~\cite{herapdf1.0}.
The QCD predictions for the structure functions are obtained by solving the DGLAP evolution equations 
at NLO (or NNLO) in the $\overline{MS}$ scheme with the renormalisation and factorisation scales chosen to be $Q^2$.
The QCD predictions for the structure functions are obtained by the convolution of the PDFs
with the NLO coefficient functions calculated using the general mass variable flavour number 
RT scheme~\cite{RTref}.
For the parametrisation of PDFs at the input scale the generic form $xf(x)=Ax^B(1-x)^C(1+Ex^2)$ is used.        
The parametrised PDFs are the gluon distribution, the valence quark distributions
and the $u$-type and $d$-type anti-quark distributions.
The normalisation parameters $A$ are constrained by the quark number and momentum sum-rules.
\begin{figure}[!h]
\center
\resizebox{0.75\columnwidth}{!}{\includegraphics{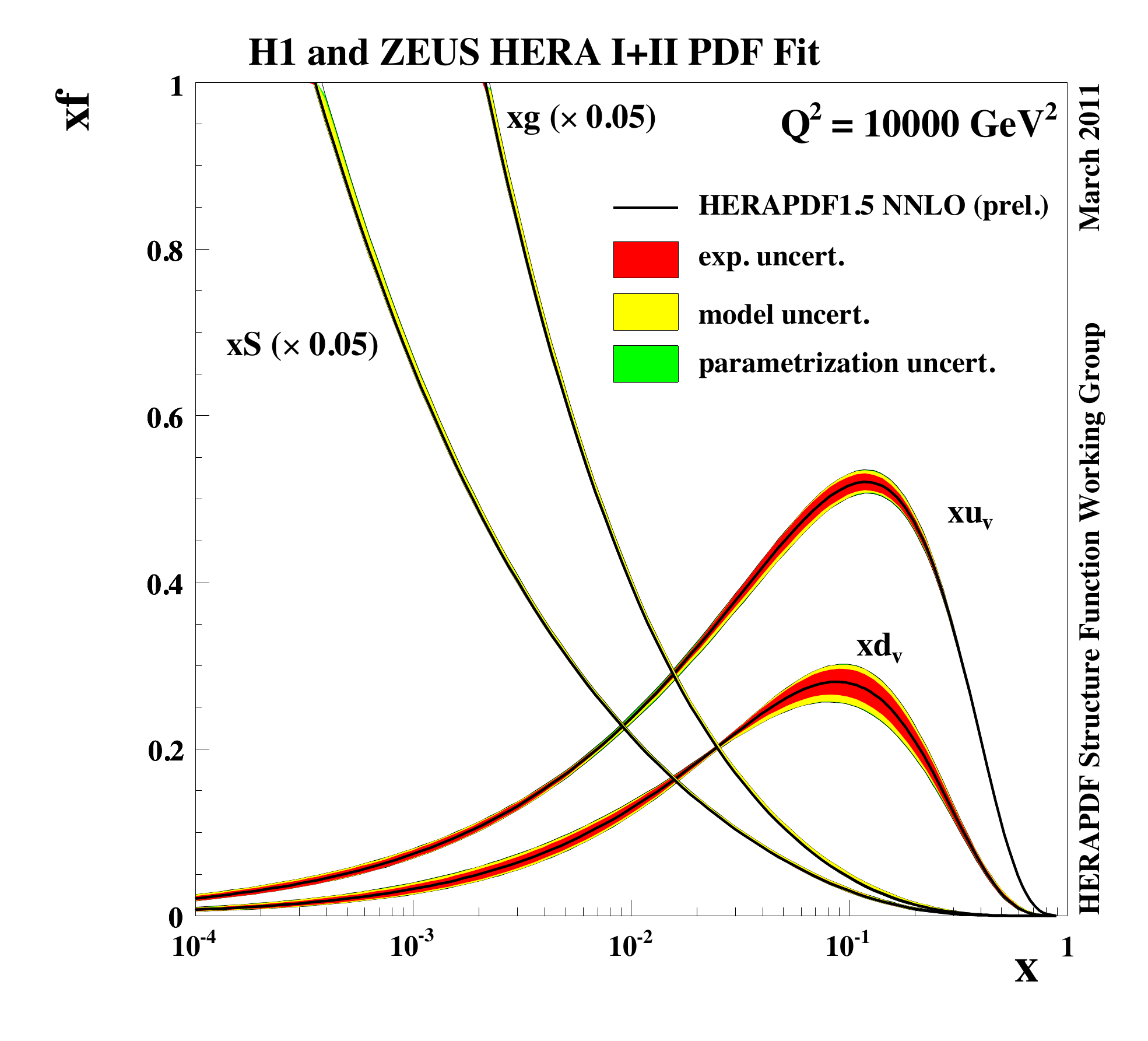}}  
 \caption{\it The parton distribution functions from HERAPDF1.5 NNLO. The gluon and sea distributions are scaled down by a factor of 20. The experimental, model and parametrisation uncertainties are shown.}
 \label{herapdf15nnlofig}
\end{figure}

In Fig. ~\ref{herapdf15nnlofig} the parton distributions HERAPDF1.5NNLO at $Q^2$ = 10000 GeV$^2$  are shown.
In addition to the experimental uncertainties, the variation of model inputs and parametrisation in the determination of HERAPDF are performed and provided as additional eigenvectors. The model uncertainties are evaluated by varying the input assumptions on minimum $Q^2$ of the data used in the fit, the stran\-ge\-ness fraction and the masses of heavy quarks. The para\-me\-tri\-sa\-tion uncertainty is formed by an envelope of the maximal deviations from 
the central fit varying parametrisation assumptions. HERA\-PDF1.5NLO and NNLO sets are the recommended HERA PDFs to be used for the predictions of processes at the LHC. The corresponding eigenvectors are available~\cite{lhapdf}.

\section{Benchmarking HERAPDF}
The PDFs are intrinsic properties of the proton and are therefore process-independent. Cross section predictions for processes in proton-(anti)proton collisions can be obtained using HERAPDF, evolved in $Q^2$ using DGLAP equations.

\begin{figure}[!ht]
\center
\resizebox{0.75\columnwidth}{!}{\includegraphics{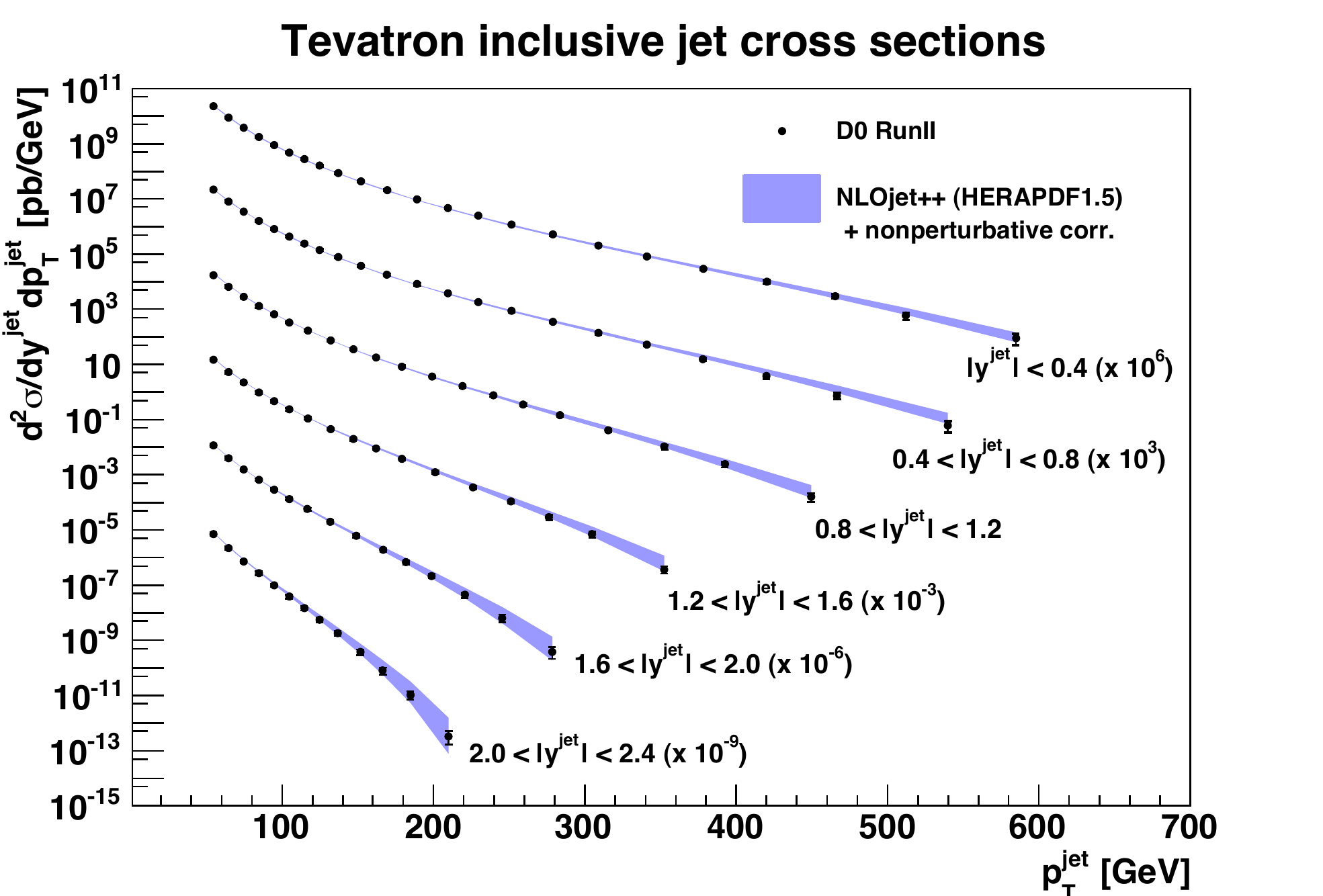}}
 \caption{\it Jet production cross section as a function of the jet transverse momentum for different ranges of pseudorapidity, as measured by the D0 collaboration. The data are represented by closed symbols. The measurement is compared to the QCD calculation at NLO based on HERAPDF1.5NLO. The total PDF uncertainty and hadronisation corrections on the prediction is shown as shaded bands. }
 \label{d0jets}
\end{figure}

\begin{figure}[!h]
 \center
  \resizebox{0.75\columnwidth}{!}{\includegraphics{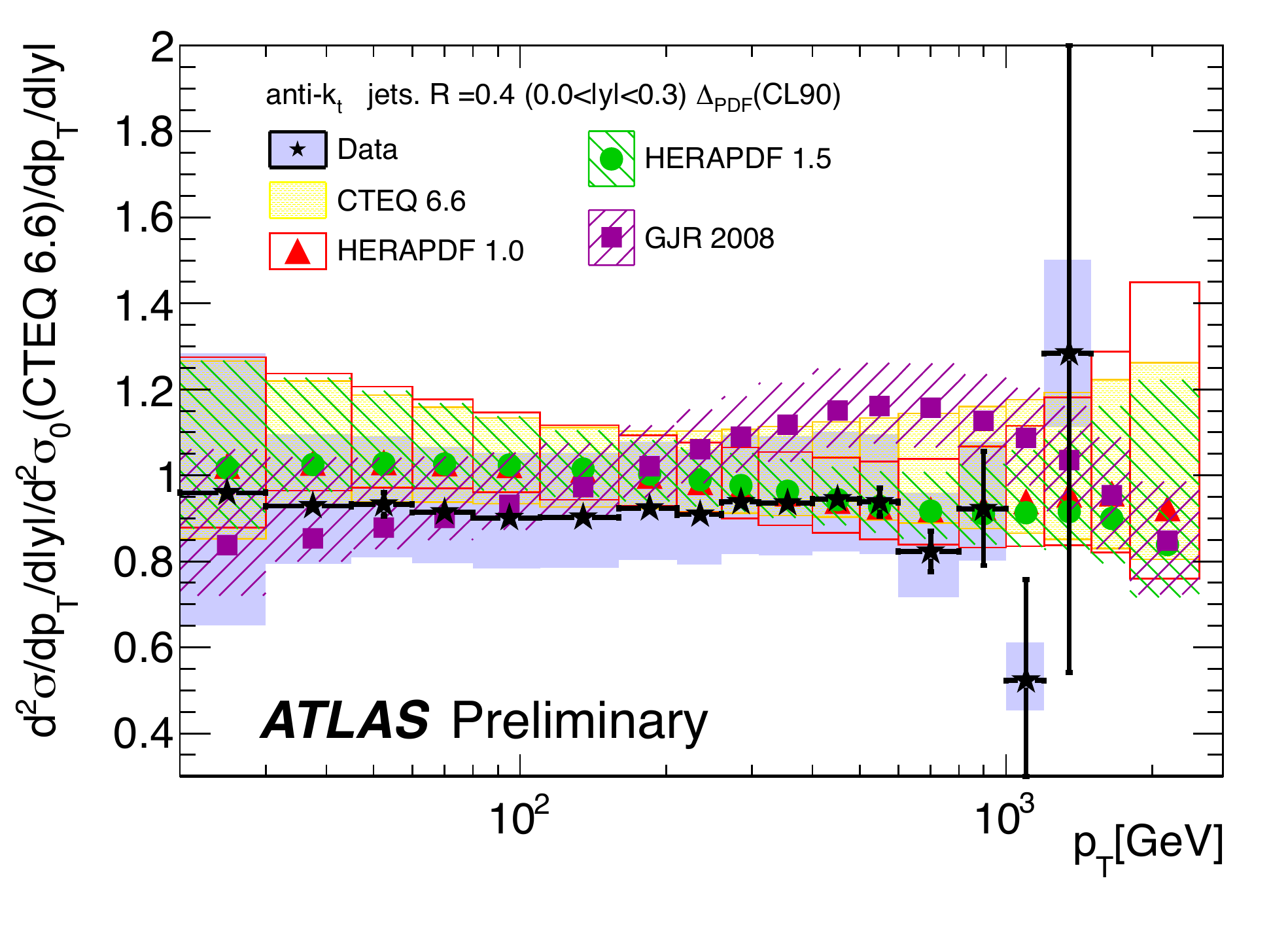}}
 \caption{\it Inclusive jet production cross section as a function of the jet transverse momentum, as measured by the ATLAS collaboration in the rapidity range $0<y<0.3$
The jets are identified using the anti-k$_t$ algorithm with $R = 0.4$. The data are represented in a ratio (stars) to the QCD prediction, using CTEQ6.6~\cite{cteq6.6} as a reference PDFs. The central value for the QCD calculation at NLO based on HERAPDF1.5NLO is represented by closed circles surrounded by the error band shown as the hashed area.}
 \label{jets_atlas}
\end{figure}

The measurements of jet production at hadron colliders is an important instrument to probe PDFs at high $x$ and also provide additional constraints on the value $\alpha_S(M_Z)$. In Fig.~\ref{d0jets} the jet production cross sections as measured by D0 experiment~\cite{d0jetpaper} is presented. The measurement is confronted with the QCD prediction at NLO~\cite{nlojet++,fastnlo}  based on HERA\-PDF1.5NLO. The data is very well described by this prediction.   
In Fig.~\ref{jets_atlas} the jet measurement from ATLAS experiment~\cite{atlas_jets} in a central rapidity bin is shown in comparison with NLO predictions using HERAPDF1.5NLO together with several other PDFs. The QCD prediction using HERAPDF1.5NLO describes the data very well.

\begin{figure}[!h]
\center
\vspace*{-0.5cm}
   \resizebox{0.75\columnwidth}{!}{\includegraphics{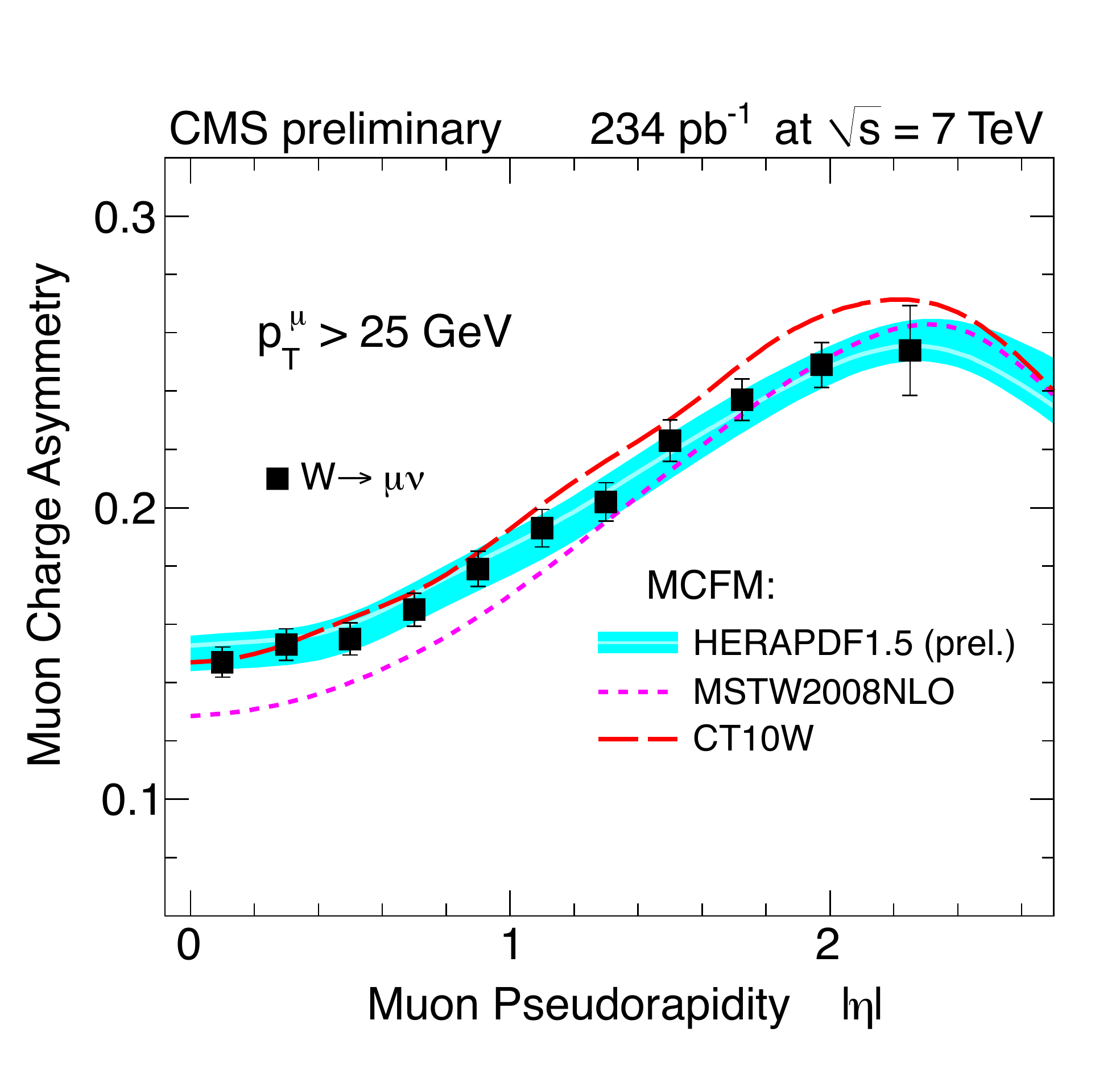}}
 \caption{\it The W muon charge asymmetry as measured by the CMS experiment. The measurement (closed symbols) is compared to the NLO prediction~\cite{mcfm} using HERAPDF1.5NLO (shaded band), MSTW08NLO (dotted line) and CT10W (dashed line).}
 \label{cms_w}
\end{figure}

Production of electroweak bosons provides important constraints on the light quark distributions. For example, the $W$ lepton charge asymmetry 
$A_l(W) = \frac{(\sigma_{W^+} - \sigma_{W^-})}{ (\sigma_{W^+} + \sigma_{W^-})} \approx \frac{(u_v - d_v)}{u_v + d_v + 2u_{sea}}$ is sensitive to the valence $u$ and $d$ quark ratio. 
The $W$-boson muon asymmetry as measured by the CMS experiment~\cite{cms_w} is shown in Fig.~\ref{cms_w}. 
The measurement is compared to NLO predictions~\cite{mcfm} obtained using HERA\-PDF1.5NLO, MSTW08~\cite{mstw08} and CT10W\cite{ct10w} PDFs. The prediction based on HERAPDF1.5NLO describes the data well.

\begin{figure}[!h]
\center
\vspace*{-0.5cm}
  \resizebox{0.75\columnwidth}{!}{\includegraphics{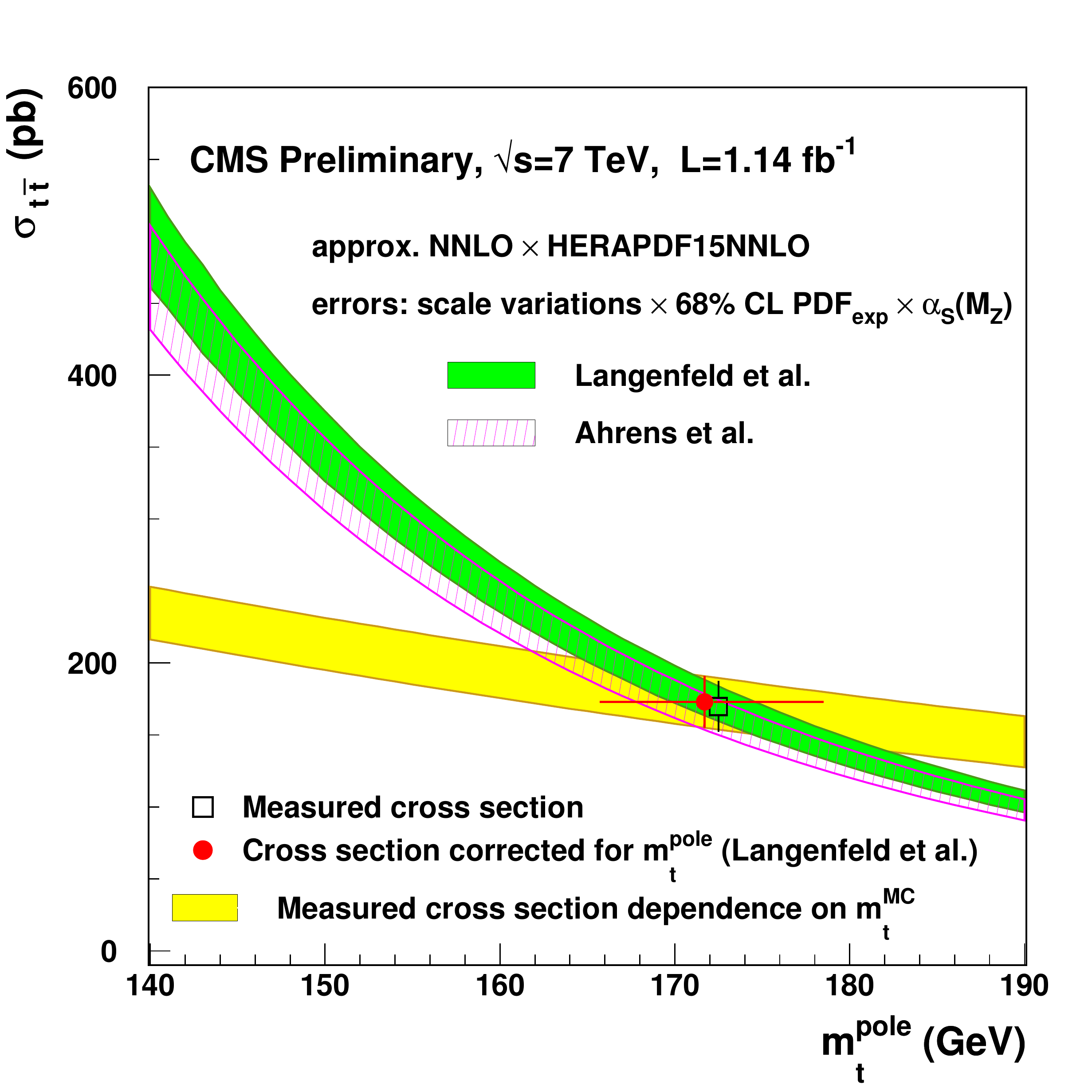}}
 \caption{\it The top-pair production cross section measured by the CMS experiment (closed square) shown at the assumption on the top mass, used in the analysis. The mass dependence of the $t\bar{t}$ cross section according to approximate NNLO QCD predictions~\cite{hathor} and ~\cite{ahrens} are represented by the shaded and hashed band, respectively. The dependence of the experimental measurement on the assumption on the $m_t$ in the simulation used for efficiency and detector corrections is shown in by light shaded band. The closed circle represents the cross section measurement, corrected for the top pole mass, extracted using the calculation~\cite{hathor}.}
 \label{top_cms}
\end{figure}

Top quark pair production at the LHC probes the gluon density at high $x$. In Fig.~\ref{top_cms} the cross-section measurement of top pair production is shown as a function of the top-quark pole mass in comparison to approximate NNLO calculations~\cite{hathor,ahrens} based on HERAPDF1.5NNLO. The theory uncertainty accounts for the variation of the QCD scales, PDFs error and the variation of $\alpha_S(M_Z)$ in the PDF. For the PDF uncertainty of HERAPDF1.5NNLO, only the eigenvectors for experimental errors are used. The predictions describe the data very well.

\section{Global benchmarking excercise}
\begin{figure}[!ht]
   \center
\resizebox{0.75\columnwidth}{!}{\includegraphics{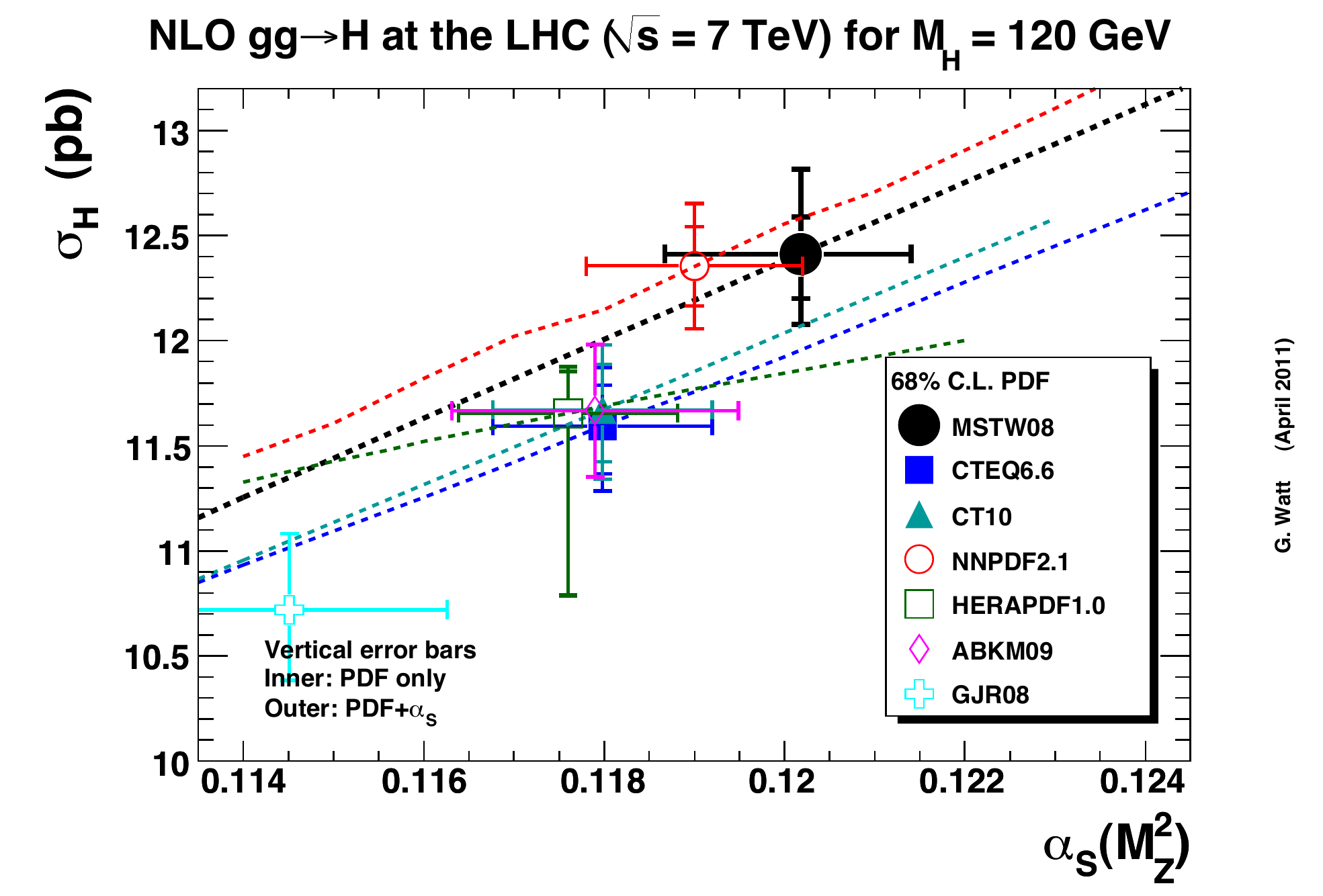}}
 \caption{\it NLO Higgs cross section predictions ($M_H = 120$ GeV) using different PDFs at the LHC
              with $\sqrt s = 7$ TeV.}
 \label{higgsfig}
\end{figure}

Presently, the determination of PDFs is carried out by several groups, namely MSTW~\cite{MSTW}, CTEQ~\cite{CTEQ}, 
NNPDF~\cite{NNPDF}, HERA\-PDF~\cite{herapdf1.0}, AB(K)M~\cite{ABKM} and GJR~\cite{GJR}.
The large number of PDF parameters and their treatment in the fitting procedure within the different groups results in differences of the PDFs provided.
In order to study these differences, a benchmarking exercise is being carried out by the PDF4LHC working group~\cite{pdf4lhc}  formed by
the members of the PDF fitting groups mentioned above. As an example,  the NLO prediction for the Higgs cross section ($M_H = 120$~GeV) at the LHC is shown 
in Fig.~\ref{higgsfig} for different PDF sets as a function of $\alpha_S(M_Z)$.  For different PDF groups not only the value of $\alpha_S (M_Z)$, but also the running of the strong coupling is different, resulting in different cross section predictions. 
The HERAPDF is an active participant in the benchmarking exercise. In contrast to other PDF groups, HERAPDF is not restricted to one particular heavy flavour treatment scheme, several schemes are implemented and can be tested. Also, by providing the PDF eigenvectors for model parameter and parametrization variations, HERAPDF allows for tests of specific parameterisation and model assumptions during the QCD ana\-ly\-sis of different data sets. In the following, the inclusion of semi-inclusive DIS data in the QCD analysis HERAPDF and the impact of these data on assumptions on $\alpha_S(M_Z)$ and the charm quark mass value in the PDF fit is discussed. 
\section{Semi-inclusive data in HERAPDFs}
Semi-inclusive measurements in DIS like jet and heavy flavour production, provide additional constraints on the PDFs when included into the QCD analysis together with inclusive DIS data. The jet production is directly sensitive to both the gluon distribution in the proton and the strong coupling $\alpha_S$. Therefore, including the jet data in the QCD analysis can help disentangling the effects from the gluon and  $\alpha_S$ in the PDF fit. Similarly, charm and beauty production in $ep$ collisions provide direct access to the gluon distribution in the proton, which also depends on the assumption of the charm and beauty mass values used in the PDF fit.
\subsection{Including jet data in the PDF fit: HERAPDF1.6}
\begin{figure}[!ht]
\hspace*{-0.4cm}
   \resizebox{0.51\columnwidth}{!}{\includegraphics{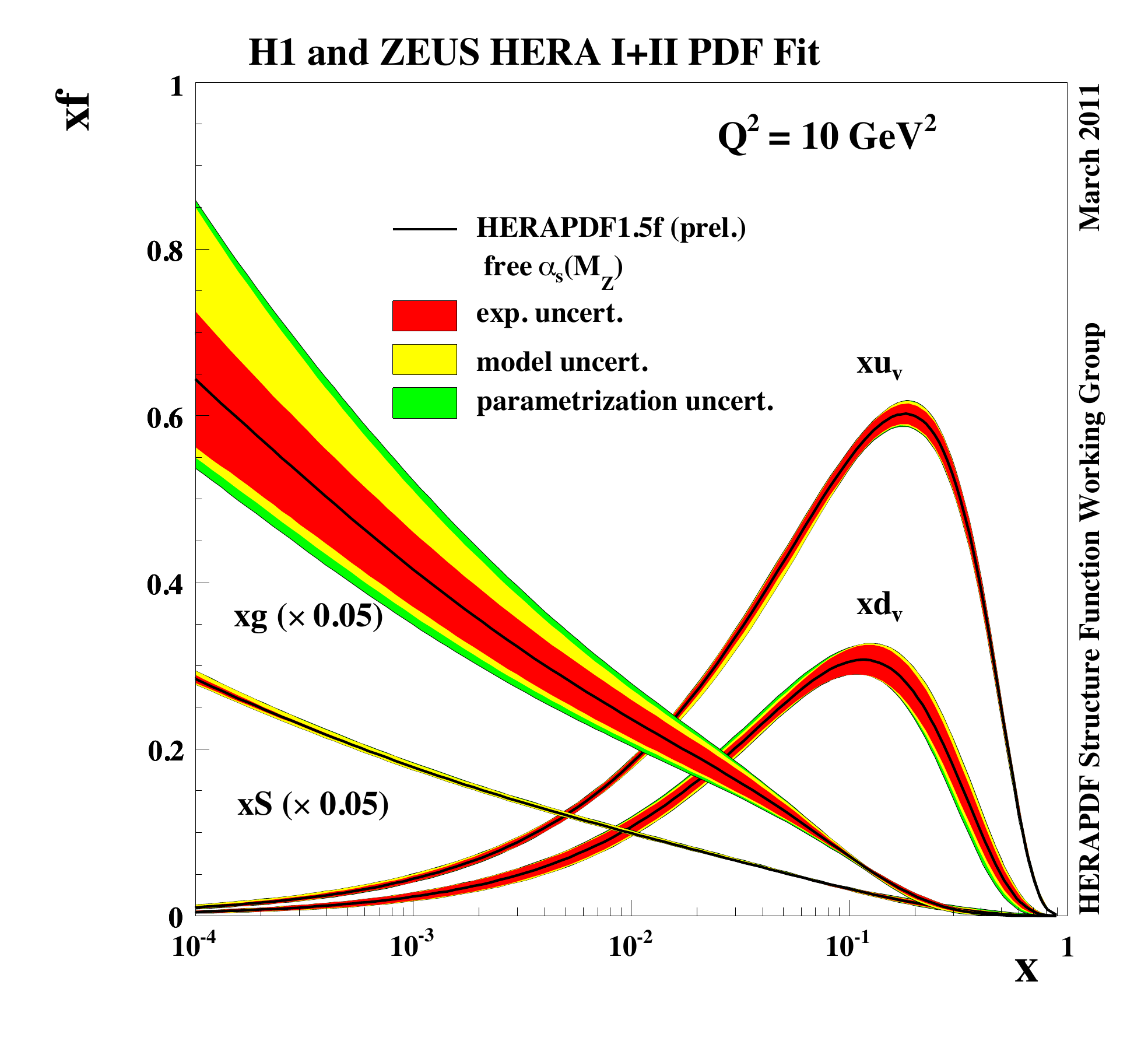}}
   \resizebox{0.51\columnwidth}{!}{\includegraphics{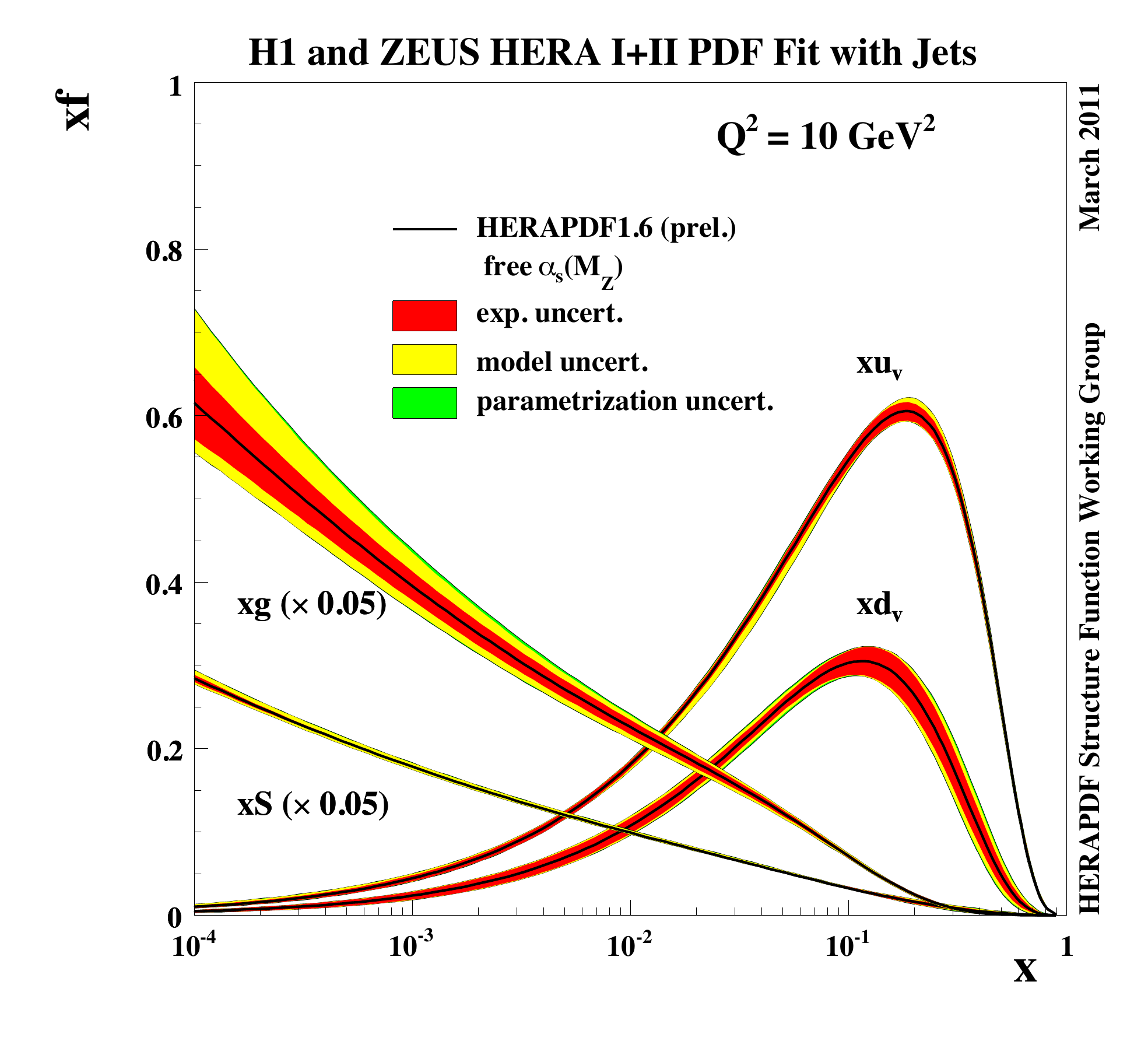}}
 \caption{\it Left panel: The parton distribution functions from HERAPDF1.5.
Right panel: the parton distribution functions from HERAPDF1.6 (with HERA jet data included in the fit). In both cases, the QCD analysis is performet treating $\alpha_s (M_Z)$ as a free parameter in the fit. The PDFs are presented for $Q^2$=10 GeV$^2$.}
 \label{pdf_free_als}
\end{figure}

\begin{figure}[!h]
\center
    \resizebox{0.7\columnwidth}{!}{\includegraphics{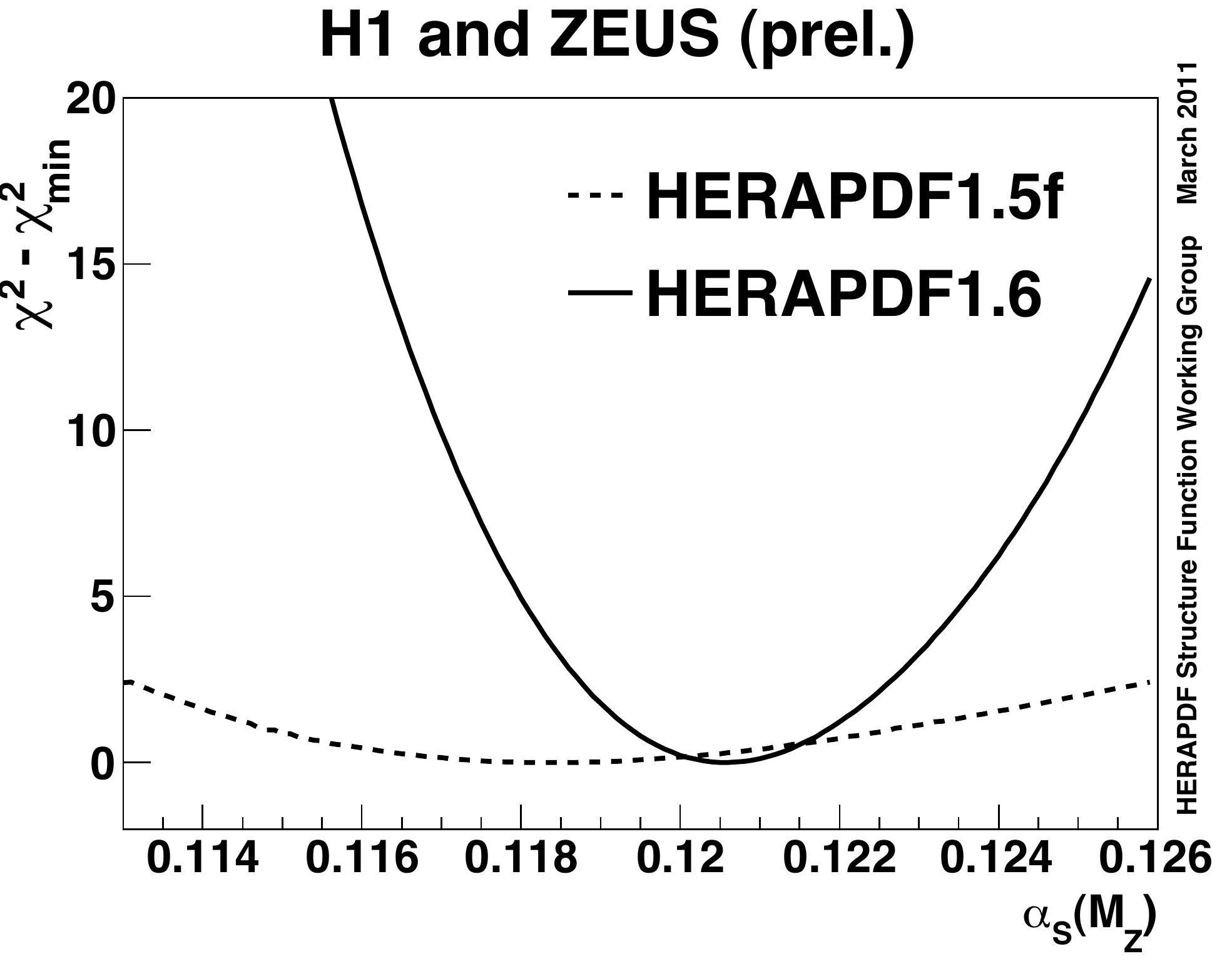}}
 \caption{\it Distribution of $\chi^2$ for PDF fit as a function of the assumption on the $\alpha_S(M_Z)$ value. The dashed line corresponds to HERAPDF1.5, where only inclusive DIS data are used. The solid line represents HERAPDF1.6, where the jet data are included.}
 \label{alfas_scan}
\end{figure}

In addition to the combined HERA inclusive DIS data as used in the QCD analysis HERAPDF1.5, H1 and ZEUS measurements of jet production cross sections~\cite{jet_h1_zeus}  are included in the PDF fit.  The resulting parton distributions HERAPDF1.6~\cite{herapdf1.6} are determined using a fixed value of $\alpha_S(M_Z)$ and also using $\alpha_S(M_Z)$ as a free parameter in the fit. 
The impact of the inclusion of jet data in the PDF fit on the gluon distribution and the value of $\alpha_S$ is demonstrated in Fig.~\ref{pdf_free_als}. Here, the PDFs obtained using the inclusive data only (HERAPDF1.5) and the PDFs resulting from including the jet data (HERAPDF1.6) are determined using $\alpha_S(M_Z)$ as a free parameter in the QCD analysis. In case of the simultaneous fit of PDFs and $\alpha_s$ in HERAPDF1.5, the uncertainties on the gluon PDF becomes large at low $x$ but as soon as the jet data are included, the correlation between the gluon PDF and $\alpha_s (M_Z)$ is reduced, resulting in significantly reduced uncertainties on the gluon PDF.
In Fig.~\ref{alfas_scan} the quality of the PDF fit in terms of $\chi^2$ is represented as a function of the assumption on the value of $\alpha_S (M_Z)$. In case of HERAPDF1.5, where only inclusive data are used, a very shallow minimum in the $\chi^2$ distribution is observed. The inclusion of the jet measurements in the fit results in the clear minimum, which allows the simultaneous determination of the PDFs and $\alpha_S(M_Z)$.

\begin{figure}[!h]
\center
    \resizebox{0.75\columnwidth}{!}{\includegraphics{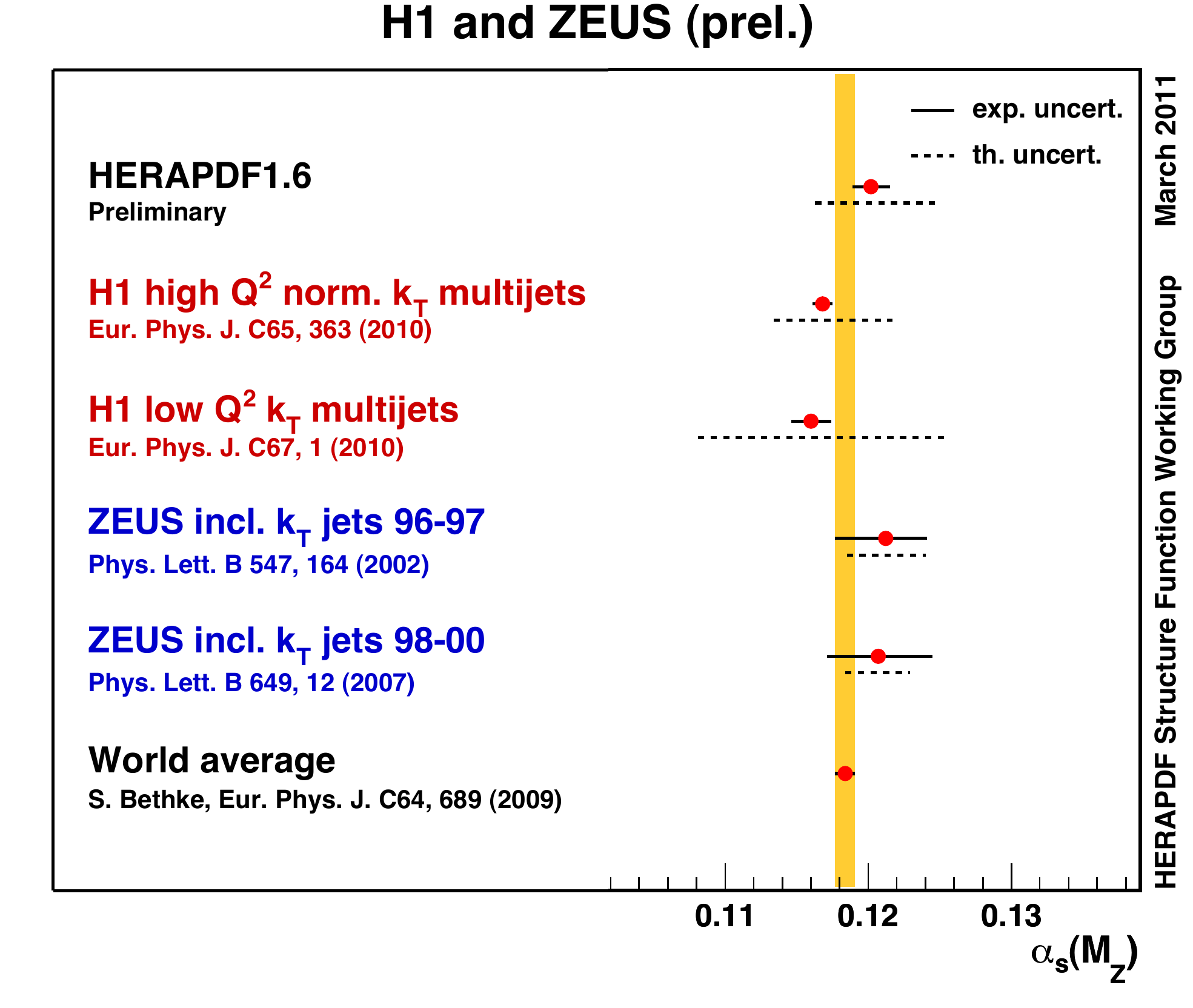}}
 \caption{\it The summary of $\alpha_S(M_Z)$ determination results using the jet production at HERA as compared to the world average. The upper point corresponds to the simultaneous determination of $\alpha_S(M_Z)$ and the PDF, as described in the text. The experimental uncertainties are represented by solid lines, the theory uncertainties are shown by dashed lines.}
 \label{alfas_all}
\end{figure}

A value of $\alpha_s (M_Z)~=~0.1202~\pm~0.0013$(exp)~$\pm~0.0007$ (mod/param) $\pm 0.0012$(hadronisation)$^{+0.0045}_{-0.0036}$(scale)
is determined~\cite{herapdf1.6}. This result is in very good agreement with different results of $\alpha_S$ determination at HERA and with the world average as shown in Fig.~\ref{alfas_all}. It is important to note, that the dominant uncertainty arises from the variation of the renormalization and factorisation scales in the NLO calculation for the jet cross sections.  This variation is used to mimic the effect of the missing contribution from higher orders. 
\subsection{Charm quark measurements in the PDF fit.}
\begin{figure}[!h]
\center
\vspace*{-0.5cm}
    \resizebox{0.75\columnwidth}{!}{\includegraphics{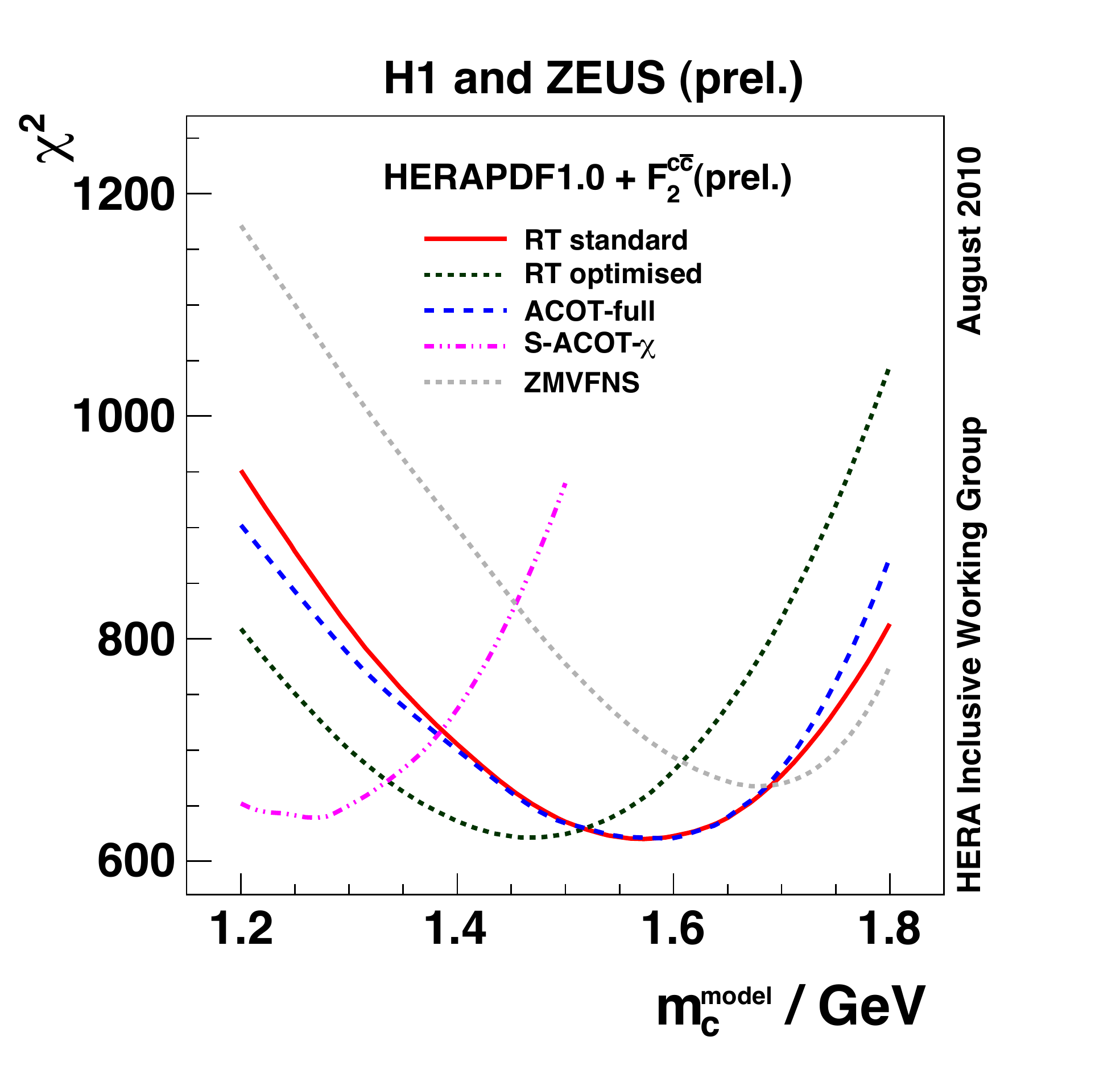}}
 \caption{\it Comparison of the $\chi^2$ distributions of fits to the inclusive HERA I + F$_2^{\bar cc}$ data using different heavy flavour schemes represented as lines of different styles.}
 \label{charm_mass_scan}
\end{figure}

The factorization scheme, used for the PDF determination depends on the assumption on the number of flavours in the proton, which varies depending on the value of the scale, which has to be compared to the threshold, at which charm and beauty quarks can be treated as partons. 
This threshold is determined by the mass of charm and beauty quarks. Therefore, the treatment of heavy quarks and the assumptions on their masses have 
particular importance in the QCD analysis of the proton structure. Different approaches to treat heavy quark (heavy quark schemes) are used by different PDF fitting groups, corresponding to different treatment of mass terms in perturbative calculations, but also implying differences in the interpretation and assumptions on the values of 
the heavy quark masses.  Measurements of charm and beauty production can help constraining some of these assumptions. The charm contribution, $F_2^c$, to the proton structure function $F_2$ is measured at H1 and ZEUS using different charm tagging techniques. These measurements are combined~\cite{hera_f2c} taking into account the correlations of the systematic uncertainties. The combined $F_2^c$ data are included in the QCD analysis of the inclusive DIS cross 
sections, and the effect on the PDFs using different assumptions on the charm quark mass, $m_c$, is studied~\cite{qcd_charm}. The sensitivity of the 
PDF fit to the $m_c$ value when using combined  $F_2^c$ is used to constrain the assumptions on $m_c$ in different heavy quark schemes~\cite{charm_mass_scan}. The $\chi^2$ values of the PDF fit including the charm data are determined as a function of  the input values of charm quark mass, $m_c^{mod}$, using different heavy quark schemes, as shown in Fig.~\ref{charm_mass_scan}.

\begin{figure}[!h]
\center
\vspace*{-0.8cm}
       \resizebox{0.75\columnwidth}{!}{\includegraphics{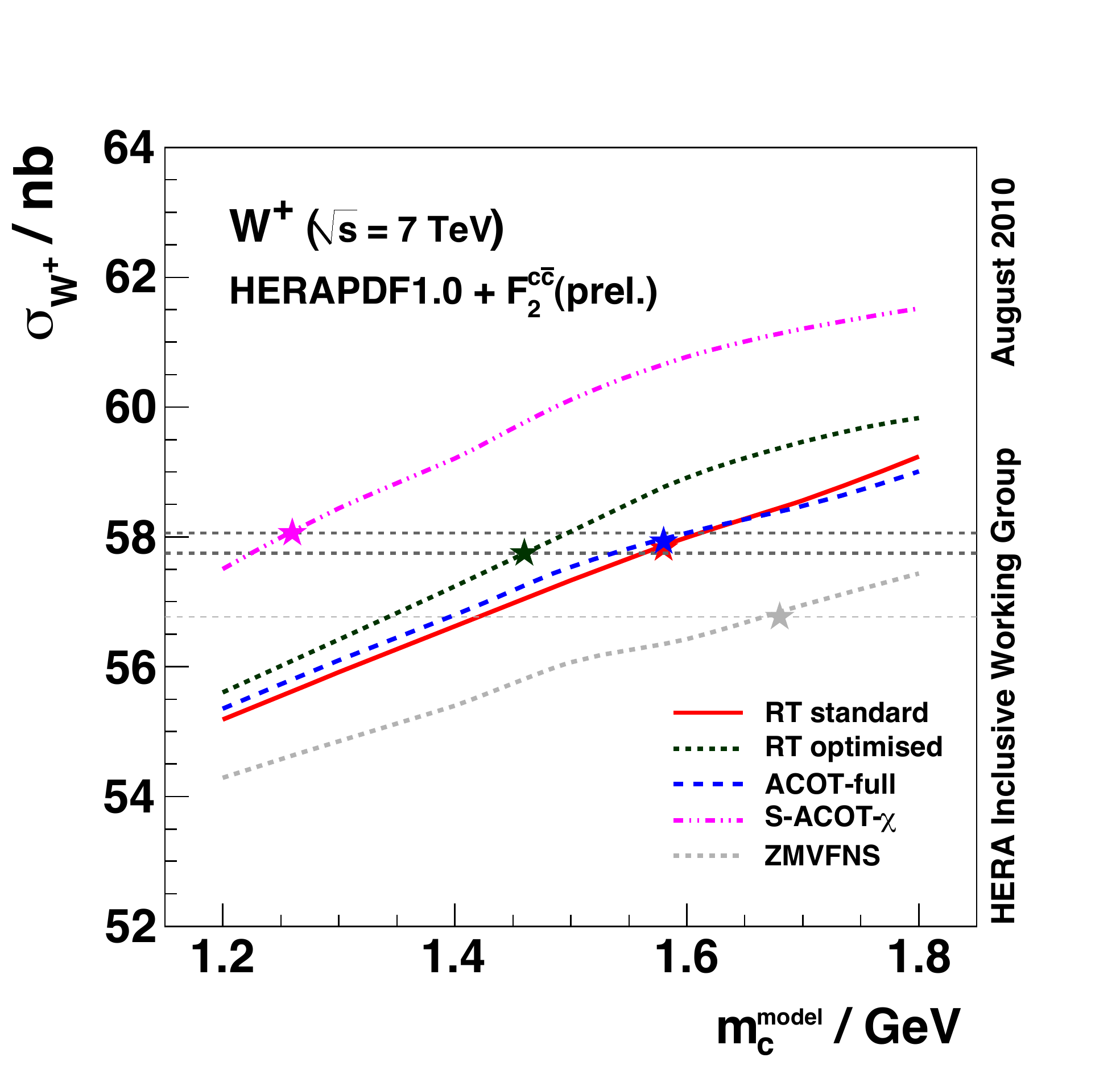}}
 \caption{\it NLO prediction of $\sigma_{W^{+}}$ at the LHC for $\sqrt s = $ 7 TeV as a function of $m_c^{mod}$ in the input PDF. The lines show predictions 
 for different VFN schemes. The stars show the predictions obtained with the optimal value of $m_c^{mod}$ used in a given scheme. The dashed horizontal 
 lines indicate the range of $\sigma_{W^{+}}$, determined for  $m_c^{mod}$ = $m_c^{mod}$ (opt).} 
 \label{charm_for_lhc}
\end{figure}

Different assumptions on $m_c^{mod}$ in VFN schemes impact the charm contribution to the sea quark distribution and thus affect the composition 
of $x \overline U(x)$ from the $x\overline u(x)$ and the $x\overline c(x)$ contributions. These in turn influence the value of the $W^{\pm}$ and $Z$ cross section 
predictions at LHC. In Fig.~\ref{charm_for_lhc} the NLO prediction~\cite{mcfm} for the $W^+$ production cross section is shown, using parton distributions evaluated with different assumptions on $m_c^{mod}$ in various heavy quark schemes. 
Taking into account the whole spread of cross section predictions using the studied schemes, an uncertainty of 7\% on the $W^+$ production cross section arises due to assumption on $m_c^{mod}$ in the PDF. However, when using the optimal values, $m_c^{mod}$ (opt), corresponding to minima from Fig.~\ref{charm_mass_scan} as constrained by HERA charm data, this uncertainty is reduced to 1\%.

\section{Summary}
Precision of the parton distribution functions is essential for accurate predictions of cross sections of the procecces at hadron colliders.    
The proton PDFs are determined using the experimental data of DIS and proton-proton collisions. Combined data of HERA 
collider experiments provide the most precise constraint on the PDFs at small and medium $x$. HERAPDF is one of the modern 
QCD analyses in which PDFs are determined. The advantages of these PDFs is no need for nuclear corrections (in contrast to PDFs 
using the fixed target data), consistent treatment of the systematic uncertainties of the experimental data and implementation of several
 phenomenological approaches of heavy flavour treatment. Currently, HERAPDF1.5 at NLO and NNLO are among the recommended 
 parton densities for predictions of LHC cross sections. Recent developments in the HERAPDF fits include the QCD analyses of HERA inclusive DIS data together with jet and charm measurements. The inclusion of the jet measurements in the HERAPDF analysis reduces the correlation between the gluon distribution and the strong coupling constant. In such a fit, the PDF is determined together with the $\alpha_S(M_Z)$ value. The resulting $\alpha_S(M_Z)$ value is in a very good agreement with the world average and its precision is limited by the missing NNLO calculation for jet production. The inclusion of the charm data reduces the correlation between the gluon density and the value of the charm mass used in different schemes of heavy flavour treatment in the PDF fit. In particular, a proper choice of the charm quark mass value is important for accurate QCD predictions of W and Z boson rates at the LHC. The QCD predictions based 
on the HERAPDF1.5  describe the measurements at Tevatron and the LHC very well. With increasing precision of the LHC data, particular 
processes like $W$-boson, jet or top-pair production will provide additional constraints on the PDFs.

The open source code for QCD analysis of different data sets HE\-RA\-Fit\-ter~\cite{herafitter} is released by the H1 and ZEUS collaborations. The program aims for implementation of all available schemes for heavy flavour treatment. HERA\-Fit\-ter is used in the ATLAS and CMS experiments to study the impact the electroweak boson production, jet production and top quark production on the proton PDFs.  

\end{document}